\documentstyle[12pt,epsf,subfigure]{article}
\newcommand{\Frac}[2]{\frac{\displaystyle #1}{\displaystyle #2}}
\newcommand{\km}{$K^0$--$\overline{K^0}$ }
\newcommand{\dm}{$D^0$--$\overline{D^0}$ }

\begin{document}
\pagestyle{empty}
\begin{titlepage}

\begin{center}
\hfill FTUV/97-43\\
\hfill IFIC/97-43\\
\hfill INFNNA-IV-97/12 \\
\hfill DSFNA-IV-97/12 \\
\hfill July, 1997
\vspace{3cm} \\
{\Large  $D^0 - \overline{D^0}$ mixing  
in the weak--gauged  \vspace*{0.4cm} \\   $U(4)_L \otimes
U(4)_R$ Chiral Lagrangian Model} \\
\vspace{2cm} 
{\large G. Amor\'os$^1$, F.J. Botella$^1$, S. Noguera$^1$, J. Portol\'es$^2$} \\
\begin{itemize}
\item[1)] Departament de F\'{\i}sica Te\`orica and I.F.I.C. (Centre Mixte 
Universitat de Val\`encia--C.S.I.C.) 
E-46100 Burjassot (Val\`encia, Spain).
\item[2)] Istituto Nazionale di Fisica Nucleare (I.N.F.N.), Sezione di Napoli
and Dipartamento di Scienze Fisiche, Universit\'a di Napoli,
I-80125 Napoli (Italy). 
\end{itemize}
\vspace{2cm}
\begin{abstract}
The \dm mixing is analyzed in a weak gauged $U(4)_L \otimes
U(4)_R$ chiral lagrangian model where the electroweak interaction
is introduced as a gauge theory over the meson degrees of freedom.
This model allows a particular realization of the G.I.M. mechanism
and then could be useful in the study of processes where G.I.M.
suppression is effective. As a test of the model we have also analyzed 
the \km mixing.
We find $\Delta m_K$ in good agreement with the experimental result and
we show that the \dm mixing is very much suppressed in agreement with
previous estimates in the Heavy Quark expansion framework.
\end{abstract}
\end{center}
\end{titlepage}
\newpage
\pagestyle{plain}
\pagenumbering{arabic}
\vspace{5cm}

\section{Introduction}
\hspace{0.5cm}Higher order effects in the perturbative treatment of the 
electroweak Standard Model (SM) play an active r\^ole both in testing
some of its fundamental ingredients and in uncovering phenomena not 
explained by the Standard framework. One of the peculiarities of the SM
is the absence (at leading order in the perturbative expansion)
of flavour changing neutral currents (FCNC). This characteristic feature 
strongly affects  flavour mixing and rare meson decays.
\par
Neglecting the tiny CP--violating amplitudes (as we do in this Letter) the
perturbative contributions to $P^0$--$\overline{P^0}$ mixing are 
straightforward to evaluate in the SM.  However the numerical 
quantification is affected by our rather
poor knowledge of some of the relevant elements of the CKM matrix and of the
quark masses.  In any case more dubious is the treatment of long--distance contributions. 
These have two main sources: a) the evaluation of the matrix element of the
four--quark $\Delta F = 2$ operator in the effective hamiltonian
(F is short for flavour), and b) the handling of mesonic intermediate states
contributing to the dispersive amplitude. Both features are characterized
by a manifestly non--perturbative origin.
\par
The relative importance of short vs. long--distance contributions
depends crucially of the flavour involved. This is so because as the flavour
is heavier the transitions involve higher transfer of momenta and therefore
non--perturbative corrections are less relevant. As a consequence 
$B_s^0$--$\overline{B_s^0}$ and $B_d^0$--$\overline{B_d^0}$ are expected
to be dominated by the perturbative regime. This is not so clear for
$K^0$--$\overline{K^0}$ and definitely not the 
case of $D^0$--$\overline{D^0}$ (due to the possible contribution of resonances in the
highly populated kinematical region of charmed mesons).

In this Letter we address the issue of non--perturbative corrections to the
mixing and therefore we will be concerned with these last two cases.
\par
At hadronic level meson mixing with $\Delta F = 2$ transitions has ushered
permanent interest because its implications~: tests of FCNC, close 
relation with CP--violation, prospects of new physics beyond the SM, etc.  
(see  \cite{GL74,DH84,DK85,CCV} and references therein).
Mixing occurs through radiative
corrections in the SM. The mass eigenstates are 
\begin{equation}
| P_{\,_1^2} \rangle \, = \, \Frac{1}{\sqrt{2}} \, \left[ \,
 |P^0 \rangle \, \pm \,  |\overline{P}^0 \rangle \, \right] \; \; ,
\label{eq:p12}
\end{equation}
and the mixing produces $P^0 \leftrightarrow \overline{P^0}$ oscillations
of amplitude $\exp (i \Delta M t)$ with 
$\Delta M = m_2 - m_1 - i ( \Gamma_2 - \Gamma_1 )/2$. The parameter
controlling the oscillation is $x \equiv \Delta m / \Gamma$ \footnote{
We call $x_K \equiv \Delta m_K / \Gamma_S$ and $x_D \equiv \Delta m_D / 
\Gamma_D$.} and consequently the mass difference between the 
eigenstates. With our normalization this mass difference is related with
the $\Delta F = 2$ transition through
\begin{equation}
\Delta m_P \, = \, \Frac{1}{m_P} \, Re \langle P^0 | {\cal H} | 
\overline{P}^{0} \rangle \; \; \; , 
\label{eq:deltam}
\end{equation}
where $m_P$ is short for the relevant pseudoscalar mass.
\par
\km and \dm mixings have notable different features in the SM. Both 
mixings have a complementary interest because test FCNC in the upper and 
lower sector of the families, respectively. Experimentally \cite{PD96}
$x_K \simeq 0.5$ and $|x_D| < 0.09$ and, as we will shortly see, 
the perturbative evaluation in the SM is able to point out these 
different behaviours because it predicts 
$\Delta m_K \sim m_c$ and $\Delta m_D \sim m_s$.  Moreover  the perturbative
evaluation gives the right order for the measured \km mixing while it 
gives a very small 
\dm mixing. This is the case because GIM mechanism \cite{GI70} is much more effective in the
lower sector of the families than in the upper one due to the quark
mass differences.
\par
Dispersive amplitudes due to intermediate mesonic states could give 
relevant non--perturbative contributions. These have been considered by 
several authors \cite{DH84,DG84,WO85,DG86,BG91,GE92} and the conclusion 
is that they indeed are important. In 
\dm could even give a result several orders of magnitude bigger than the
perturbative one. This might be due to the fact that the expected 
GIM suppression could be 
spoiled by the bad $SU(3)$ breaking observed in several decays of
charmed mesons \cite{DG86,KA95}.
\par 
We have proposed a weak-gauged chiral model up to and including charmed
mesons in order to implement at hadronic level all the structure of symmetries
of the SM, specifically including the GIM mechanism \cite{PO91}. Therefore, 
the $SU(3)_L \otimes SU(3)_R$ chiral symmetry of the strong interactions 
must be extended to a $U(4)_L \otimes U(4)_R$ symmetry in order to 
implement the weak $SU(2)_L$ representations. Nevertheless it must be 
clear that our model is not pretended to be a model for the strong 
interaction. Of course chiral symmetry is badly broken in the charm sector. 
We expect that this
model may provide a realistic approach to those processes where relevant
features of the flavour sector of the SM like GIM play a significant r\^ole.
In this Letter we evaluate in our model \cite{AM96}, the leading order 
result for the \km and \dm mixings. 

In Section 2 we remind briefly the perturbative evaluations of the mixings in 
the SM. An outline of the basic ingredients of our model is sketched in 
Section 3. Then in Sections 4 and 5 we specify our results for \km and \dm, 
respectively. We end with our conclusions in Section 6.

\section{The perturbative regime}
\hspace*{0.5cm} The short--distance contribution to meson mixing in the SM
is represented by the usual box diagrams. In order to carry out this
calculation one starts with the weak effective hamiltonian 
${\cal H}_{eff}^{\Delta F = 2}$ in terms of four--quark operators. We explain 
both \km and \dm mixings in turn.

\subsection{\km}
\hspace*{0.5cm} The effective hamiltonian (we are not interested in 
CP-violation) including the next-to-leading logarithmic QCD corrections, 
is given by
\cite{IL81,GW83}
\begin{equation}
{\cal H}_{eff}^{\Delta S=2} \, = \, \Frac{1}{16 \pi^2} \, G_F^2  \, m_c^2
\, \lambda_c^2  \, \eta_1 \, S \left( \Frac{m_c^2}{M_W^2} \right) \, 
\alpha_s(\mu^2)^{(-2/9)} \, 
{\cal O}^{\Delta S = 2} \; \; , 
\label{eq:hds2}
\end{equation}
where 
\begin{equation}
{\cal O}^{\Delta S = 2} \, = \,  (\overline{s}_L ^{\alpha} \gamma_{\mu} 
d_L^{\alpha}) 
(\overline{s}_L^{\beta} \gamma^{\mu} d_L^{\beta}) \, \; \; \; , 
\label{eq:ops2}
\end{equation}
$\lambda_c = V_{cd} V_{cs}^*$ with
$V_{ij}$ the CKM matrix, $\alpha$ and $\beta$ are colour indices and
\begin{equation}
S(x) \, =  \, \left[ 1 \, + \, \Frac{9}{1-x} \, - \, 
\Frac{6}{(1-x)^2} \, - \, \Frac{6 x^2}{(1-x)^3} \ln x \, \right] \; \; .
\label{eq:sx}
\end{equation}
In Eq.~(\ref{eq:hds2}) $\eta_1$ carries the information of strong 
QCD corrections  \cite{HN95}.
\par
A part of the perturbative uncertainties in order to get a firm prediction
we need to do the evaluation of the matrix element of the four--quark
operator in Eq.~(\ref{eq:ops2}) between the asymptotic kaon states. This,
of course, involves the unsolved problem of hadronization and our ignorance is
expressed by the $B$ parameter defined as
\begin{equation}
\langle \overline{K}^0 | {\cal O}^{\Delta S = 2} | K^0 \rangle \, = \, 
\Frac{4}{3} \, f_K^2 m_K^2 \, B(\mu) \; \; \; ,
\label{eq:bpar}
\end{equation}
where, in the leading logarithmic approximation in the strong coupling,
\begin{equation}
B(\mu) \, =  \,  B_K \, \alpha_s(\mu^2)^{+2/9} \; \; \; .
\label{eq:bkll}
\end{equation}
A simple factorization approach and vacuum insertion \cite{GL74}
gives $B(\mu) = 1$.  The non-perturbative
parameter $B_K$ has been calculated using lattice gauge theories 
\cite{GG92}, 
$1/N_c$ expansion \cite{BB88}
and QCD sum rules \cite{DE86} leading to $B_K = 0.7 \pm 0.2$. 
The QCD hadron duality approach \cite{PR91} favours lower 
values $B_K = 0.4 \pm 0.1$.

\subsection{\dm}
\hspace*{0.5cm} The effective hamiltonian (neglecting the contribution
of the $b$ quark due to the suppression of the relevant CKM matrix 
elements) is given by \cite{DK85,DG86}
\begin{equation}
{\cal H}_{eff}^{\Delta C = 2} \, = \, - \, \Frac{1}{2 \pi^2} \, G_F^2 \,
\widetilde{\lambda_s} \widetilde{\lambda_d} \, \Frac{(m_s^2-m_d^2)^2}{m_c^2}
 \, \left( \, 
{\cal O}_1^{\Delta C=2} \, + \, 2 \, {\cal O}_2^{\Delta C=2} \, \right) \; \; ,
\label{eq:hdc2}
\end{equation}
where $\widetilde{\lambda_i} = V_{ci}^* V_{ui}$ and the four-quark
operators are defined as
\begin{eqnarray}
{\cal O}_1^{\Delta C = 2} & = & (\overline{u}_L^{\alpha} \gamma_{\mu} 
c_L^{\alpha}) 
(\overline{u}_L^{\beta} \gamma^{\mu} c_L^{\beta}) \; \; , \nonumber \\
& & \label{eq:opc2} \\
{\cal O}_2^{\Delta C = 2} & = & (\overline{u}_L^{\alpha} c_R^{\alpha})
 (\overline{u}_L^{\beta} c_R^{\beta}) \; 
\; . \nonumber
\end{eqnarray}
This last operator  has no analogous in the \km case and it is due to the 
non--negligible charm quark mass $m_c$ carried by two of the four external
legs of the box diagram.
\par
Analogously to the \km mixing the evaluation of the matrix elements of the
two four--quark operators involves non--perturbative aspects that, this
time, are parameterized by two quantities $B_D$ and $B'_D$ for the
two operators in Eq.~(\ref{eq:opc2}) respectively, and defined to be the
unity if factorization and vacuum insertion approximations are used.
Corrections to this value are potentially large but are not expected to
change the order of magnitude.
\par
At any rate from Eq.~(\ref{eq:hdc2}) we can conclude that \dm mixing is 
suppressed typically by a factor $ \sim m_s^4/m_c^4$ over the \km case.

\section{The model}
\hspace*{0.5cm}Our lagrangian is \cite{PO91}~:
\begin{equation}
       {\cal L} = {\cal L}_{mesons} + {\cal L}_{Higgs}
                  + {\cal L}_{HM}  + \ldots \; \; ,
\label{eq:model}
\end{equation}
where the dots are short for pure gauge boson terms.
Here ${\cal L}_{mesons}$ contains the strong interaction between mesons
and the couplings of mesons to the gauge bosons.
We have a set of $16$ scalar and $16$ pseudoscalar fields which
we assign to the
$(4,\bar{4}) \oplus (\bar{4},4)$ representation of the chiral
$U(4)_L \otimes U(4)_R$ group. We denote the meson matrix by
$U = \Sigma + i \Pi$, where $\Sigma$ is the scalar and $\Pi$
the pseudoscalar matrices of fields.
The explicit expression for the pseudoscalar matrix is

\renewcommand{\arraystretch}{0.3}
\begin{eqnarray*}
      \Pi & = & \left( \begin{array}{cccc}
                   \frac{1}{2}\eta_{\circ} + 
                   \frac{1}{\sqrt{2}}\pi^{0} + & & 
                   &  \\ & \pi^{+} & K^{+} & \overline{D^{0}} \\
                   \frac{1}{\sqrt{6}}\eta_8 + \frac{1}{\sqrt{12}}
                   \eta_{15} & & & \\
                   & & & \\ & & & \\
                    & \frac{1}{2}\eta_{\circ} -
                   \frac{1}{\sqrt{2}}\pi^{0} + & & \\
                   \pi^{-} & & K^{0} & D^{-} \\
                   & \frac{1}{\sqrt{6}}\eta_8 +
                   \frac{1}{\sqrt{12}}\eta_{15} & & \\
                   & & & \\ & & & \\
                   & & \frac{1}{2}\eta_{\circ} -
                   \frac{2}{\sqrt{6}}\eta_8 + & \\
                   K^{-} & \overline{K^{0}} & & D_S ^{-} \\
                   & & \frac{1}{\sqrt{12}}\eta_{15} & \\
                   & & & \\ & & & \\
                   D^{0} & D^{+} & D_S ^{+} &
                   \frac{1}{2}\eta_{\circ} -
                   \frac{3}{\sqrt{12}}\eta_{15}
                   \end{array}            
             \right) \; \; .
\end{eqnarray*}
\begin{equation}
         \;
\label{eq:matrix}
\end{equation}

\renewcommand{\arraystretch}{1}

A similar matrix can be written for the scalar mesons. Our notation
for scalar mesons is $\sigma_{\circ},\sigma_8, \sigma_{15}, \sigma^+,
\sigma_3, \kappa, \delta, \delta_S$ instead of $\eta_{\circ},
\eta_8,\eta_{15}, \pi^+, \pi^{0}, D $ and $D_S$ respectively.
\par
With these definitions ${\cal L}_{mesons}$ is 

\begin{equation}
      {\cal L}_{mesons} = \frac{1}{2} Tr[(D^{\mu}U')^{\dagger}
                                         (D_{\mu}U')] -
                           V_{chiral}(U) \; \; ,
\label{eq:strong}
\end{equation}
where $V_{chiral}$ is the chiral potential
\begin{eqnarray}
      V_{chiral}(U) & = & - \mu_{\circ}^2 \; Tr(U^{\dagger}U) + 
                            \nonumber \\*
                    &   &
                            \mu_{\circ}^2 \; [  a \;
                                 Tr(U^{\dagger}U)^2  + 
                                        b \; (Tr(U^{\dagger}U))^2  + 
                                        c \; (det U + det U^{\dagger}
                                            )  ] \; \; , 
\label{eq:potential}
\end{eqnarray}
with $\mu_{\circ}^2 > 0$ in order to develop spontaneous breaking
of chiral symmetry.
The covariant derivative is:
\begin{equation}
      D_{\mu} U' = \partial_{\mu} U' -
                        i g \overrightarrow{T} \cdot \overrightarrow{W_{\mu}} U' -
                        i g' Y_L B_{\mu} U' +
                        i g' U' Y_R B_{\mu} \; \; , 
\label{eq:dcov}
\end{equation}
with 
\begin{equation}
              U' = S U S^{\dagger} \; \; .
\label{eq:mix}
\end{equation}
Here $\overrightarrow{W_{\mu}}$ and $B_{\mu}$ are the gauge bosons related with
the $SU(2)_L$ and the $U(1)_Y$ groups. The $\overrightarrow{T}$ matrices are the
$SU(2)$ generators and $Y_L$ and $Y_R$ are the left and right 
hypercharges. The matrix $S$ will be the Cabibbo rotation once
the $SU(2)_L \otimes U(1)_Y$ symmetry gets spontaneously  broken.
With our definitions for $U$ we have
\begin{equation}
\begin{array}{cc}
      \vec{T} = \frac{1}{2} \left( \begin{array}{cc}
                                   \vec{\tau} & 0 \\
                                   0 & \tau^1 \vec{\tau} \tau^1
                                   \end{array} \right) \; \; , 
      &
      Y_L = \frac{1}{6} I_{4\times4} \; \; , \\    \\
      Y_R = \frac{1}{3} \left( \begin{array}{cccc}
                               2 &  &  &  \\
                                 & -1 &  &  \\
                                 &  & -1 &  \\
                                 &  &  & 2
                               \end{array} \right) \; \; , 
      &
      S = \left( \begin{array}{cccc}
                 1 & 0  & 0  & 0 \\
                 0 & \cos \theta_c & \sin \theta_c & 0 \\
                 0 & - \sin \theta_c & \cos \theta_c & 0 \\
                 0 & 0 & 0 & 1
                 \end{array} \right) \; \; ,
\end{array}
\label{eq:genera}
\end{equation}
with $\vec{\tau}$ the usual Pauli matrices and $\theta_c$ the 
Cabibbo angle. The charge operator is $Q = Y_R = Y_L + T_3$.
\par
In Eq.~(\ref{eq:model})  ${\cal L}_{Higgs}$ is the usual lagrangian 
for the minimal model
of Higgs of the Standard Theory. ${\cal L}_{HM}$, finally, is a 
Higgs--mesons coupling term which will give masses to the mesons
 after the spontaneous symmetry breaking of the weak symmetry. In
order to write in a compact form this term we introduce the 
usual Higgs doublet as a $4 \times 4$ matrix:
\begin{equation}
     H = \left( \begin{array}{cccc}
                \frac{1}{\sqrt{2}} (\psi - i \chi) &
                s^{+} & 0 & 0 \\
                - s^{-} & \frac{1}{\sqrt{2}} (\psi + i \chi) &
                0 & 0 \\
                0 & 0 & \frac{1}{\sqrt{2}} (\psi + i \chi) &
                - s^{-} \\
                0 & 0 & s^{+} &
                \frac{1}{\sqrt{2}} (\psi - i \chi)
                \end{array}  \right) \; \; \; . 
\label{eq:higgs}
\end{equation}
We consider only the simplest local gauge invariant term

\begin{equation}
     {\cal L}_{HM} = Tr (A S^{\dagger} H^{\dagger} S U + h.c.) \; \; , 
\label{eq:hm}
\end{equation}
where $S$ is given in Eq.~(\ref{eq:genera}) and the most  general form of $A$ 
assuming isospin symmetry is
\begin{equation}
     A = \left( \begin{array}{cccc}
                \alpha & & & \\
                 & \alpha & & \\
                 & & \gamma & \\
                 & & & \delta
                \end{array} \right) \; \; \; . 
\label{eq:mass}
\end{equation}
${\cal L}_{HM}$ breaks explicitly the $SU(4) \otimes SU(4)$ chiral
symmetry. Through the spontaneous breaking of the chiral and weak 
symmetry the matrices of mesons and Higgs get a non--zero vacuum
expectation value 
\begin{equation}
\begin{array}{cc}
\langle \circ | U | \circ \rangle \equiv F = \Frac{1}{\sqrt{2}}
\left( \begin{array}{cccc}
        f_{\alpha} & & & \\
        & f_{\alpha} & & \\
        & & f_{\gamma} & \\
        & & & f_{\delta} 
        \end{array} \right) , \; \; \;  &  
\langle \circ | H | \circ \rangle \equiv  \Frac{1}{\sqrt{2}} \phi_{\circ}
I_{4 \times 4} \; \; \; . 
\end{array}
\label{eq:efes}
\end{equation}
Therefore from Eq.~(\ref{eq:hm}) we get
\begin{equation}
{\cal L}_{HM} = \Frac{\phi_{\circ}}{\sqrt{2}} Tr ( A ( U + U^{\dagger})) +
Tr ( A S^{\dagger} \tilde{H}^{\dagger} S U + h.c.) \; \; , 
\label{eq:hmdef}
\end{equation}
where
\begin{equation}
\tilde{H} = H - \langle \circ | H | \circ \rangle \; \; \; . 
\label{eq:hshift}
\end{equation}
We note that the lagrangian in Eq.~(\ref{eq:hm})
 transforms under chiral $SU(4) \otimes SU(4)$ as the 
$(4, \bar{4}) \oplus (\bar{4}, 4)$ representation and, therefore,
the first term in Eq.~(\ref{eq:hmdef}) is similar to 
the usual explicit breaking of chiral symmetry.
\par
Some aspects of this model are worth to emphasize:
\begin{itemize}
\item[a)] The starting point of our ideas
is to believe that the symmetries of the Standard Theory are essential
to describe the weak processes of hadrons, and any model for them has
to support not only its symmetries but its symmetry breaking patterns too.
Our realization with two complete families allows to implement rigorously 
the structure of the SM at hadronic level.
\item[b)]The GIM mechanism is naturally implemented in our scheme.
Therefore we have not flavour changing neutral currents at leading order
in the perturbative expansion and in this way the model might presumably be 
trusted in the study of processes where GIM plays a significant r\^ole.
\item[c)]The model has been tested satisfactorily in the study of non--leptonic 
D decays into two pseudoscalars \cite{BN93a}, non--resonant $D \rightarrow
PPP$ \cite{BN93b} and also, at one--loop level, in radiative
rare kaon decays  ($K^+ \rightarrow \pi^+
\ell^+ \ell^-, K_S \rightarrow \pi^{\circ} \ell^+ \ell^-$) \cite{PO91} with 
satisfactory results.
\end{itemize}
Hence we think it is worth to use the model just described
in order to study \km and \dm mixings. We have already seen in Section 2
that the short--distance contribution to \km mixing is dominated by the
charm degree of freedom while the strange quark dominates \dm mixing. 
Our model provides these ingredients and, moreover, allows the evaluation
of long--distance dispersive contributions mediated by the pseudoscalar
(and maybe scalar) mesons.
\par
Therefore we proceed to calculate them at ${\cal O}(G_F^2)$ in the
weak perturbation expansion. Due to our ignorance on the scalar masses
and in order to present simplified analytical expressions, we first 
analyze the case in which $m_{scalar} \gg 
m_{pseudoscalar}$, that is $\mu_{\circ}^2 \rightarrow \infty$ in
Eq.~(\ref{eq:potential}) but with $\mu_{\circ}^2 c \equiv c'$ constant in 
order to keep the $\eta_{\circ}$ mass finite. In this case $F$ in 
Eq.~(\ref{eq:efes}) is
\begin{equation}
F = \Frac{1}{\sqrt{2}} f_{\circ}  \left[I_{4 \times 4} \, + \, 
{\cal O} \left( \Frac{m_{Pseudoscalar}^2}{\mu_{\circ}^2} \right) \right] \; \; , 
\label{eq:fexp}
\end{equation}
with
\begin{equation}
f_{\circ}^2 = \Frac{1}{a + 4 b} \; \; \; . 
\label{eq:f0}
\end{equation}
With our normalization $f_{\circ} \, = \, f_{\pi} \, \sim 90 MeV$ in the
chiral limit.
\par
Using this expansion, at leading order, we get a relation between
the pseudoscalar masses, 
\begin{equation}
m_{D_S}^2 \, - \, m_D^2 \, = \, m_K^2 \, - \, m_{\pi}^2 \; \; . 
\label{eq:massre}
\end{equation}
Eq.~(\ref{eq:fexp}) is a crude approximation for the charm sector, we must 
remember here that $U(4)_L \otimes U(4)_R$ is not a symmetry for the strong 
interaction in nature. Therefore we have also considered the case of finite 
scalar masses in both calculations where one goes beyond the rough 
approximation represented by Eq.~(\ref{eq:fexp}) and (\ref{eq:massre}).
 
We use the renormalizable $R_{\xi}$ gauge \cite{FL72} where there is 
no mixing 
meson--W and instead we have a meson--charged Higgs and meson mixings as
shown in Fig. 1.
\par
At this order the two different topologies of Feynman diagram 
contributions to the mixing are those in Fig. 1. Both diagrams
are separately finite but only the sum of the two is gauge independent.
\par

\section{\km mixing}
\hspace*{0.5cm} In order to test the model we can calculate $\Delta m_K$.
 The problem of evaluating long--distance contributions
to the \km mixing has been addressed extensively in the bibliography
\cite{DG84,BG91,DG82,BS84}. These are due to intermediate hadronic
states in the oscillation and are presumably dominated by the octet
of pseudoscalars: $K^0 \leftrightarrow (\pi^0 , \eta) \leftrightarrow
\overline{K^0}$, $K^0 \leftrightarrow \pi \pi \leftrightarrow 
\overline{K^0}$, etc. An estimate of these contributions gives results
comparable with the experimental mass difference. However the evaluation
has significant uncertainties.

%%%%%%%%%%%%%%%%%%%%%%%%%%%%%%%%%%%%%%%%%%%%%%%%%%%%%%%%%%%%%%%%%%%%%%
\epsfverbosetrue
\begin{figure}
\epsfxsize=14cm
\centering
\epsffile{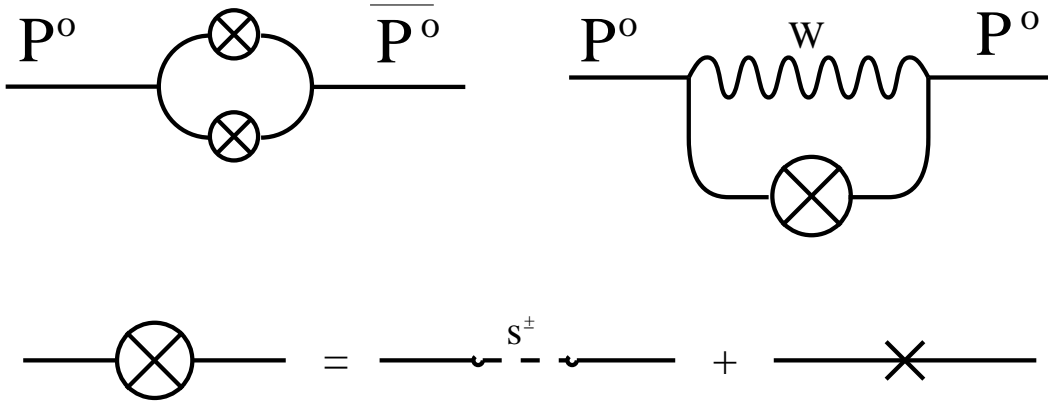}
\caption{The loop diagrams are the contribution of the model 
to the \km and \dm mixing. The crossed propagator is 
the sum of the Higgs $s^{\pm}$ contribution plus the meson mixing. 
  }
\end{figure}

%%%%%%%%%%%%%%%%%%%%%%%%%%%%%%%%%%%%%%%%%%%%%%%%%%%%%%%%%%%%%%%%%%%%%%

\par
The computation of the \km mixing in our model through the Feynman 
diagrams of Fig. 1 gives for the difference of masses:
\begin{eqnarray*}
\Delta m_K &  =  & 
\Frac{ G_F^2  \sin^2 \theta \cos^2 \theta f_{\pi}^2}{8 \pi^2 m_K} \,
\left[ \, (m_D^2 \, - \, m_{\pi}^2)^2 \, + \, (m_D^2 \, - \, m_{\pi}^2)
( \, m_{D_S}^2 \ln m_{D_S}^2 \, - \, m_K^2 \ln m_K^2 \, ) \right. 
 \\  
& & \; \; \; \; \; \; \; \; \; \; \; \; \; \; \; \; \; \; \; \; \;
\left. + \, m_D^2 (m_D^2 \, + \, m_K^2) \left( \,
\Frac{m_{D_S}^2}{m_D^2 - m_{D_S}^2} \ln  \Frac{m_D^2}{m_{D_S}^2} 
\, - \, \Frac{m_K^2}{m_D^2 - m_K^2} \ln  \Frac{m_D^2}{m_K^2}  \, 
\right) \right.  \\ 
& &  \; \; \; \; \; \; \;  \; \; \; \; \; \; \; \; \; \; \; \; \; \; \; \;
\left. + \, m_{\pi}^2 (m_{\pi}^2 \, + \, m_K^2) \left( \, 
\Frac{m_K^2}{m_{\pi}^2 - m_K^2} \ln \Frac{m_{\pi}^2}{m_K^2} \, - \, 
\Frac{m_{D_S}^2}{m_{\pi}^2 - m_{D_S}^2} \ln  \Frac{m_{\pi}^2}{m_{D_S}^2}
\, \right) \right.  \\  
& & \; \; \; \; \; \; \; \; \; \; \; \; \; \; \; \; \; \; \; \;
\left. - \, \Frac{1}{2}(m_K^2 \, - \, m_{D_S}^2)^2 P(m_{D_S}^2,m_{D_S}^2)
\, - \, \Frac{1}{2} (m_K^2 \, - \, m_D^2)^2 P(m_D^2, m_D^2) \right.
 \\
& & \; \; \; \; \; \; \; \; \; \;
\left. - \, \Frac{1}{2}(m_K^2 \, - \, m_{\pi}^2)^2 P(m_{\pi}^2,
m_{\pi}^2) \, + \, 
(m_K^2 \, - \, m_D^2)(m_K^2 \, - \, m_{\pi}^2) P(m_D^2, m_{\pi}^2)
\right]  ,
\end{eqnarray*}
\begin{equation}
\label{eq:mkmod}
\end{equation}

where $P(m_1^2, m_2^2)$ is defined through
\begin{equation}
I_2(m_1^2, m_2^2) \, = \, - \, \Frac{i}{16 \pi^2} \, 
\left\{ \, \Frac{2}{D-4} \, + \, \gamma \, - \, 
\ln (4\pi) \, - \, 2 \, + \, P(m_1^2,m_2^2) \, \right\} \; \; ,
\label{eq:scatwo}
\end{equation}
and $I_2(m_1^2,m_2^2)$ the scalar two--point function.
\par
Several comments are in order about our result Eq.~(\ref{eq:mkmod}):
\begin{itemize}
\item[-]Our result is finite and only depends on the masses of the
pseudoscalar mesons and the meson decay constants once the mass 
expansion induced by the assumption $m_{scalar} \gg m_{pseudoscalar}$
is carried out.
\item[-]We can easily see the effect of the GIM cancellation (in the
$SU(4)_F$ limit $m_u = m_c$) by inputting $m_D = m_{\pi}$ and
$m_K = m_{D_S}$ in Eq.~(\ref{eq:mkmod}) that gives evidently
$\Delta m_K = 0$.
\item[-]From the observation that the short--distance contribution to 
$\Delta m_K$ scales with $m_c^2$ in Eq.~(\ref{eq:hds2}) one should expect 
in our model that our result for $\Delta m_K$ might scale like
$m_D^4$. However using the mass relation Eq.~(\ref{eq:massre}) in 
Eq.~(\ref{eq:mkmod}) we see that the term in $m_D^4$ cancels. This 
fact provides a further suppression beyond GIM mechanism.
\end{itemize}

Using the mass relation in Eq.~(\ref{eq:massre}) and expanding the
logarithms in turn the result in Eq.~(\ref{eq:mkmod}) can be written as
\begin{equation}
\Delta m_K \, = \, \Frac{1}{4 \pi^2} G_F^2 \, \sin^2 \theta \, 
\cos^2 \theta \, f_\pi^2 \, \Frac{19}{12} \, (1 + \omega_K) \, m_D^2 \, m_K \; \; \; .
\label{eq:definem}
\end{equation}
where $\omega_K = {\cal O} \left( \Frac{m_{K,\pi}^2}{m_D^2} \right)$ 
and our result gives
$\omega_K \simeq -0.09$ providing a tiny $10 \%$ correction to the
leading term. Using the values of 
the decay constants and masses of \cite{PD96}
we get
\begin{equation}
\Delta m_K \,  =  \, 3.45 \, \times \, 10^{-15} \, GeV \; \; \; .
\label{eq:reskm} 
\end{equation}
Inputting the experimental result for $\Gamma_S$ \cite{PD96} we predict 
$x_K = 0.49$. Our results are in rather
good agreement with the experimental figure \cite{PD96}
\begin{equation}
\Delta m_K^{exp} \,  =  \, (3.491 \, \pm \, 0.009) \times \, 10^{-15} 
\, GeV \; \; \; . 
\label{eq:expkm} 
\end{equation}

This fact can be understood because the dispersive part of the amplitude 
is dominated by the $\Delta I=3/2$ transitions and, therefore, through the 
$\langle \pi^+ \pi^{\circ}| {\cal H}_W | K^+ \rangle $ amplitude \cite{DG82} 
that is well recovered in our model. 

One could also consider the corrections coming from finite 
scalar masses. These happen to depend only on two parameters~:
$f_{\gamma}/f_{\alpha}$ defined in 
Eq.~(\ref{eq:efes}) and $c'$ related to the masses of the mesons 
$\eta$ and $\eta'$. This dependence is showed in Fig. 2.
Inside reasonable values of these parameters our 
prediction $\Delta m_K$ in Eq.~(\ref{eq:reskm}) could 
only be modified in a factor 2. We can conclude that the model gives 
reasonable values for $\Delta m_K$, in the range 
$( 1.8 - 3.5) \times  \, 10^{-15} \, GeV$.

\section{\dm mixing}
\hspace*{0.5cm}As has been noted previously in Section 2 the 
short--distance contribution to the \dm mixing is very much 
suppressed relatively to the \km mixing, roughly a factor 
$m_s^4/m_c^4$. A careful evaluation shows that the suppression is not
that much but in any case the perturbative contribution gives
\begin{equation}
\Delta m_D^{box} \, \simeq \, 10^{-2} \, \Delta m_K^{box}
\; \; \; .
\label{eq:compdel}
\end{equation}
The study of long--distance contributions to \dm mixing has been
considered previously \cite{WO85,DG86,GE92} with very different
conclusions. While Wolfenstein \cite{WO85}, Donoghue et al.
\cite{DG86} and Kaeding \cite{KA95} conclude that 
the dispersive non--perturbative
contribution to \dm mixing might be huge at the level of providing
$\Delta m_D \simeq \Delta m_K$ due to the fact that there is 
a strong $SU(3)$ breaking that overcomes the GIM suppression, Georgi 
\cite{GE92} using arguments
of a Heavy Quark expansion concludes that this is unlikely, and a 
careful study in this framework \cite{OR93} seems to confirm
this assessment.
\par
The evaluation of $\Delta m_D$ at leading order in our model goes
again through the corresponding Feynman diagrams in Fig. 1. By 
expanding in $m_{Pseudoscalar}/m_{Scalar}$ and keeping the leading
term we get
\begin{eqnarray*}
\Delta m_D & = & 
\Frac{G_F^2  \sin^2 \theta \cos^2 \theta f_{\pi}^2 }{8 \pi^2 m_D}  
\left[ \, (m_K^2 \, - \, m_{\pi}^2  )^2 \, + \,
(m_K^2 \,  - \, m_{\pi}^2)( m_{D_S}^2 \ln m_{D_S}^2 \, - \, 
m_D^2 \ln m_D^2) \right. \\
& & \; \; \; \; \; \; \; \; \; \; \; \; \; \; \; \; \; \; \;
\left. + \, m_K^2 \, ( m_K^2 \, + \, m_D^2) \, 
\left( \Frac{m_{D_S}^2}{m_K^2 - m_{D_S}^2} \, \ln \Frac{m_K^2}{m_{D_S}^2}
\, - \, \Frac{m_D^2}{m_K^2 - m_D^2} \, \ln  \Frac{m_K^2}{m_D^2} \right)
\right. \\
& & \; \; \; \; \; \; \; \; \; \; \; \; \; \; \; \; \; \; \; \; \; \;
\left. + \, m_{\pi}^2 \, ( m_{\pi}^2 \, + \, m_D^2) \, 
\left( \Frac{m_{D}^2}{m_{\pi}^2 - m_{D}^2} \, \ln \Frac{m_{\pi}^2}{m_{D}^2}
\, - \, \Frac{m_{D_S}^2}{m_{\pi}^2 - m_{D_S}^2} \, \ln 
 \Frac{m_{\pi}^2}{m_{D_S}^2} \right)
\right. \\
& &  \; \; \; \; \; \; \; \; \; \; \; \; \; \; \; \; \; \; \;
\left. - \, \Frac{1}{2} (m_D^2 \, - \, m_{D_S}^2)^2 \, P(m_{D_S}^2,
m_{D_S}^2) \, -  \, \Frac{1}{2} (m_D^2 \, - \, m_K^2)^2 \, 
P(m_K^2,m_K^2) \, \right. \\
& &  \; \; \; \; \; \; \; \; \; \;
\left. - \, \Frac{1}{2} (m_D^2 \, - \, m_{\pi}^2)^2 \, 
P(m_{\pi}^2,m_{\pi}^2) \, + \, (m_D^2 \, - \, m_K^2)(m_D^2 - m_{\pi}^2)
\, P(m_K^2,m_{\pi}^2) \, \right] ,
\end{eqnarray*}
\begin{equation}
\; \label{eq:mdmod}
\end{equation}
where $P(m_1^2,m_2^2)$ has been defined in Eq.~(\ref{eq:scatwo}). Again
we find a finite result in terms of the pseudoscalar masses and the meson
decay constants. Moreover GIM suppression (for exact $SU(3)_F$ symmetry,
$m_d = m_s$) is translated
now into the meson language as $m_K = m_{\pi}$ and $m_{D_S} = m_D$ that
gives $\Delta m_D = 0$. 
\par
Using the mass relation in
Eq.~(\ref{eq:massre}) and expanding over the charmed meson mass we have:
\begin{equation}
\Delta m_D \, = \, \Frac{1}{4 \pi^2} G_F^2 \, \sin^2 \theta \, 
\cos^2 \theta \, f_\pi^2 \,  \left( 1 \, - \, \Frac{\pi}{4 \sqrt{3}} \right) 
(1 \, + \, \omega_D)  \; \Frac{m_K^4}{m_D} \; \; \; ,
\label{eq:mdsimp}
\end{equation}
with $\omega_D \simeq 0.05$ carrying the contribution of higher order
in $m_{K,\pi}^2 / m_D^2$.
As we see the $\Delta m_D$ scales with the 
 mass of the kaon to the fourth over the mass of the charmed meson. We 
notice that our result implies a suppression
\begin{equation}
\Delta m_D \, \sim \, \Frac{m_K^3}{m_D^3} \, 
\Delta m_K \;  . 
\label{eq:delresup}
\end{equation}
From Eq.~(\ref{eq:mdsimp}) we predict
\begin{equation}
\Delta m_D \, = \, 2.21 \, \times \, 10^{-17} \, GeV \; \; \; \; ,
\label{eq:dmdmod}
\end{equation}
and with the experimental $\Gamma_D$ \cite{PD96} we have 
$x_D \, \simeq \, 1.4 \, \times \, 10^{-5} \;$ .
There is only an experimental upper limit on $\Delta m_D$ \cite{PD96}
\begin{equation}
|\Delta m_D^{exp} | \, < \, 1.38 \times 10^{-13} \, GeV \; \; \; \; \; 
\; \; (90 \% CL) \; \; \;,
\label{eq:dmdexp}
\end{equation}
that is four orders of magnitude bigger than our result.

%%%%%%%%%%%%%%%%%%%%%%%%%%%%%%%%%%%%%%%%%%%%%%%%%%%%%%%%%%%%%%%%%%%%%%
\epsfverbosetrue
\begin{figure}
\begin{center}
\leavevmode
\vspace*{-1cm}
\hspace*{-1cm}
\hbox{%
\epsfxsize=18cm
\epsffile{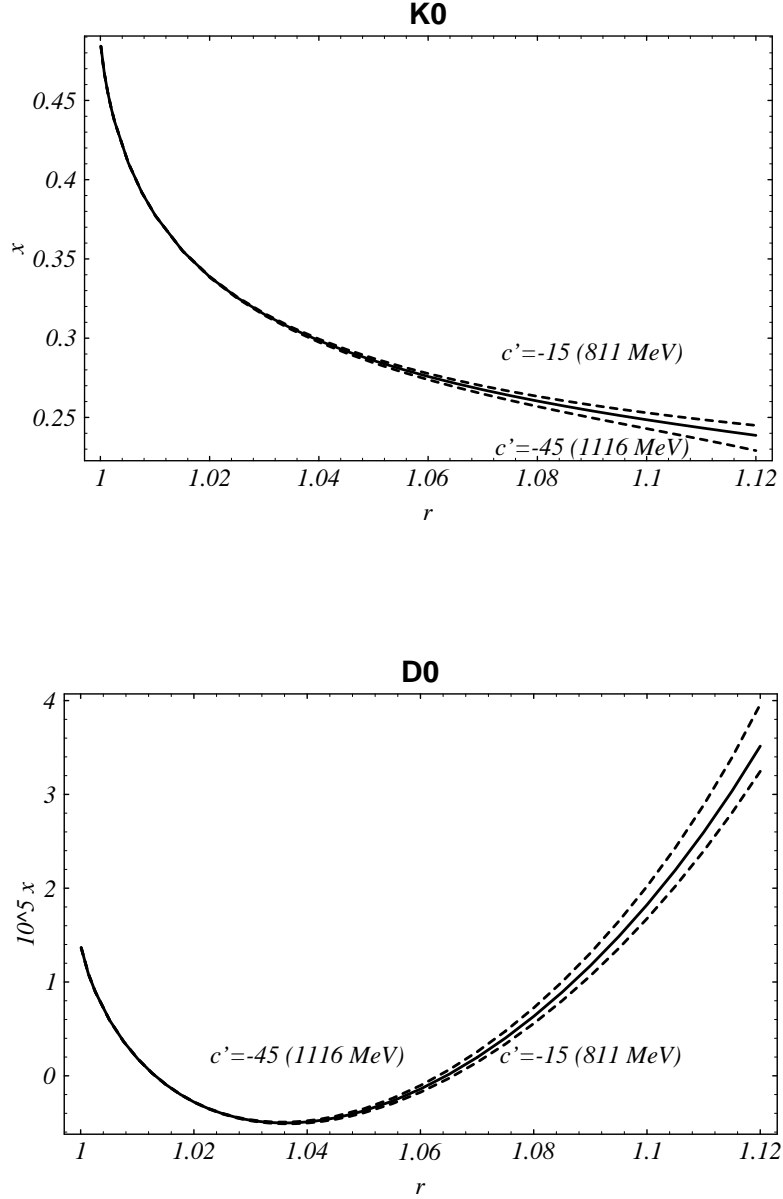}}
\end{center}
\vskip-4cm
\caption{ $x = \Delta m_P / \Gamma$ as function of $r=f_\gamma/f_\alpha$. This 
ratio is related to the mass of the scalar mesons. The left-hand side ($r=1$) 
of the plots 
correspond to the limit $m_{scalars} \longrightarrow \infty$ and the 
right-hand side to $m_{scalars} \sim 2 \; GeV$. The parameter $c'=
-15,-30,-45$ is related to the mass of the  $\eta'$ meson; we indicate the 
corresponding mass in brackets.}
\end{figure}

%%%%%%%%%%%%%%%%%%%%%%%%%%%%%%%%%%%%%%%%%%%%%%%%%%%%%%%%%%%%%%%%%%%%%%

\par
However,
there are many dynamical effects in the energy region of $1-2 \, GeV$
which could influence the final states in $D$ decays and then
modify our prediction.
Analogously to the \km mixing we can give an estimate of the dependence
of $\Delta m_D$ on the masses of the scalar mesons (Fig. 2). In this case the
dependence is bigger than in $\Delta m_K$ but for reasonable values of 
$f_{\gamma}/f_{\alpha}$ and $c'$ there is no change 
in the order of magnitude estimate. Moreover we notice that the 
sign of $\Delta m_D$ could be changed due to the contribution of the
scalar mesons. We think that, in the SM, a bigger order of magnitude
for the \dm mixing than our result implied by Eq.~(\ref{eq:dmdmod}) 
could only be expected through a large $SU(3)_F$ breaking possibly
induced by final--state interactions generated through 
vector meson resonances. This effect we know it is important
in non--leptonic decays of charmed mesons \cite{BS87} and therefore
could be relevant in the \dm mixing. However, it can be expected  
a cancellation in the total contribution of the vector mesons due to 
the GIM mechanism. 
Actually we expect, from these new intermediate channels, 
contributions without changing noticeably the order 
of magnitude. Once incorporated all the contributions, the final--state 
interactions correspond to a unitarity transformation and then it is not 
expected a modification in the mixing.
 
\par
A result for $\Delta m_D$ as given by our prediction in 
Eq.~(\ref{eq:dmdmod}), if true, is out of the experimental reach
in the foreseen future. However an experimental result at level
of $\Delta m_D > 10^{-16} \, GeV$ could be a signal of new effects
outside of the SM. An overview of the experimental techniques
and problems has been given in \cite{LI95}. The recent experimental
determination of the doubly Cabibbo suppressed channel 
$D^0 \rightarrow K^+ \pi^-$ \cite{PD96}, even still with a big error
to reduce \footnote{In \cite{BN93a} we predicted, with the
model used here, this width and our result $\Gamma (D^0 \rightarrow
K^+ \pi^-) = 3.3 \times 10^{-16} GeV$ is in very good agreement 
with the experimental figure $\Gamma (D^0 \rightarrow K^+ \pi^-) 
= (4.6\pm2.2) \times 10^{-16} GeV$.}, represents a good step forward
in this direction.

\section{Conclusion}
\hspace*{0.5cm}
We have studied the \km and \dm mixings in a weak gauged 
$U(4)_L \otimes U(4)_R$ chiral lagrangian model that incorporates
the GIM mechanism at hadronic level. We have got 
a result for $\Delta m_K$ in rather good agreement with the
experimental result and a prediction for $\Delta m_D$ that shows
a strong GIM cancellation and provides 
$x_D$ in the range $ (4,-1) \times \, 10^{-5}$ . 
Our result is in good agreement with
previous estimates in the framework of a Heavy Quark expansion 
\cite{GE92,OR93}. Other calculations giving largest $ \Delta m_D$ \cite{DG86} 
are not really in contradiction with our result. Reevaluating the estimate in 
\cite{DG86} with the actual experimental values for the widths we obtain more 
cancellation than the originally assumed in \cite{DG86}. Using theoretical 
results for the widths from \cite{BN93a} and Eq.~(\ref{eq:mdmod}) we can also 
observe that the phenomenological expression used in \cite{DG86} must be 
corrected, giving additional cancellations. From the present analysis and 
previous works on the subject we can safely conclude that a value of 
$\mid \! x_D \! \mid$ 
over $10^{-4}$ would be a clear signal of new physics in the charm sector.

\vspace*{1cm}
\noindent
{\bf Acknowledgements.}
\noindent
J.P. wishes to thank G. D'Ambrosio and G. Isidori for useful conversations on 
the topic of this paper.
This work has been partially supported by CICYT under grant AEN96-1718 and 
DGES under grant PB95-1096. G.A. is also supported by a fellowship from 
Generalitat Valenciana.

%\par
%\newpage

%\hspace*{-0.6cm}{\large \bf Figure Captions}
%\\
%\\
%{\bf Figure 1}~: $\;$ The loop diagrams are the contribution of the model 
%to the \km and \dm mixing. The crossed propagator is 
%the sum of the Higgs $s^{\pm}$ contribution plus the meson mixing. 
%\\
%\\
%{\bf Figure 2}~: $\;$ $x = \Delta m_P / \Gamma$ as function of
% $r=f_\gamma/f_\alpha$. This 
%ratio is related to the mass of the scalar mesons. The left-hand side ($r=1$) 
%of the plots 
%correspond to the limit $m_{scalars} \longrightarrow \infty$ and the 
%right-hand side to $m_{scalars} \sim 2 \; GeV$. The parameter $c'=
%-15,-30,-45$ is related to the mass of the  $\eta'$ meson; we indicate the 
%corresponding mass in brackets.

\end{document}